\documentclass[11pt]{article}
\usepackage{moriond1,epsfig}

\usepackage{amssymb}
\usepackage{graphicx}
\usepackage{bm}
\usepackage{citesort}
\usepackage{makeidx}
\usepackage{multicol}
\usepackage{bbm}
\usepackage{amsmath}
\usepackage{epsfig}
\long\def\comment#1{ }

\newcommand{\beq}{\begin{eqnarray}}
\newcommand{\eeq}{\end{eqnarray}}
\newcommand{\be}{\begin{eqnarray}}
\newcommand{\ee}{\end{eqnarray}}

\newcommand{\Lam}{\Lambda_{{\rm QCD}}}

\newcommand{\lan}{\langle}
\newcommand{\ran}{\rangle}

\def\simge{\mathrel{%
   \rlap{\raise 0.511ex \hbox{$>$}}{\lower 0.511ex \hbox{$\sim$}}}}
\def\simle{\mathrel{
   \rlap{\raise 0.511ex \hbox{$<$}}{\lower 0.511ex \hbox{$\sim$}}}}
\def\bigs{\mathrel{
   \rlap{\raise 0.531ex \hbox{$>$}}{\lower 0.531ex \hbox{$<$}}}}

\def\empile#1\over#2{\mathrel{\mathop{\kern 0pt#1}\limits_{#2}}}

\begin{document}

\vspace*{2.5cm}
\title{FROM STATISTICAL PHYSICS TO HIGH ENERGY QCD}

\author{Edmond IANCU\footnote{Presented at 40th Rencontres 
de Moriond on QCD and High Energy Hadronic Interactions, 
La Thuile, Aosta Valley, Italy, 12-19 Mar 2005.}
}

\address{Service de Physique Th\'eorique, CEA/DSM/SPhT,
CE Saclay, F-91191 Gif-sur-Yvette, France}

\maketitle\abstracts{I discuss recent progress in understanding
the high--energy evolution in QCD, which points towards a
remarkable correspondence with the reaction--diffusion problem of
statistical physics.}



Recently, there has been significant progress in our understanding
of QCD at high energy, based on the observations that \texttt{(i)}
the gluon number fluctuations play an important role in the
evolution towards saturation and the unitarity limit
\cite{MS04,IMM04} and \texttt{(ii)} the QCD evolution in the
presence of fluctuations and saturation is in the same
universality class as a series of problems in statistical physics,
the prototype of which being the `reaction--diffusion' problem
\cite{MP03,IMM04,IT04}. These observations have developed into a
rich correspondence between high--energy QCD and modern problems
in statistical physics, which relates topics of current research
in both fields, and which has already allowed us to deduce some
insightful results in QCD by properly translating the
corresponding results from statistical physics
\cite{MP03,IMM04,IT04}.

To put such theoretical developments into a specific physical
context, let us consider $\gamma^*$--proton deep inelastic
scattering (DIS) at high energy, or small Bjorken--$x$. We shall
view this process in a special frame in which most of the total
energy is carried by the proton, whose wavefunction is therefore
highly evolved, while the virtual photon has just enough energy to
dissociate long before splitting into a quark--antiquark pair in a
colorless state (a `color dipole'), which then scatters off the
gluon distribution in the proton. The transverse size $r$ of the
dipole is controlled by the virtuality $Q^2$ of $\gamma^*$
(roughly, $r^2\sim 1/Q^2$), so for $Q^2\gg \Lam^2$ one can treat
the dipole scattering in perturbation theory. But for sufficiently
small $x$, even such a small dipole can see a high--density
gluonic system, and thus undergo strong scattering.

Specifically, the small--$x$ gluons to which couple the projectile
form a {\em color glass condensate\,}\cite{CGC} (CGC), i.e., a
multigluonic state which is characterized by high quantum
occupancy, of order $1/\alpha_s$, for transverse momenta $k_\perp$
below the {\em saturation momentum} $Q_s(x)$, but which becomes
rapidly dilute when increasing $k_\perp$ above $Q_s$. The
saturation scale rises very fast with the energy, $Q_s^2(x)\sim
x^{-\lambda}$, and is the fundamental scale in QCD at high energy.
In particular, the external dipole is strongly absorbed provided
its size $r$ is large on the scale set by $1/Q_s$, whereas for
$r\ll 1/Q_s$ one rather has weak scattering, or `color
transparency'.

\begin{figure}
\begin{center}
\centerline{\epsfig{file=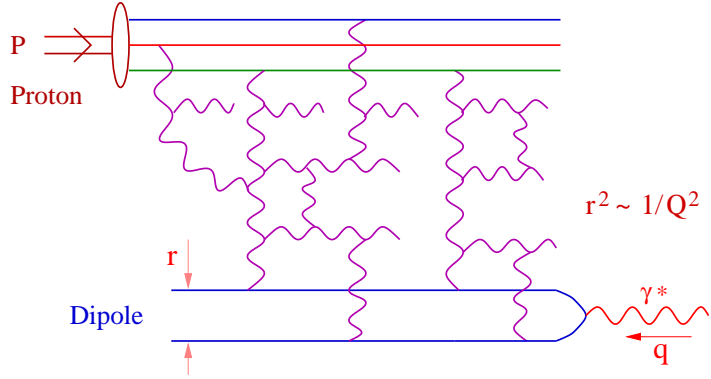,height=5.cm}} \caption{\sl
An instantaneous gluon configuration in the proton wavefunction as
`seen' in DIS at small $x$.
    \label{dis_blois1}}
\end{center}
\end{figure}

\begin{figure}[t]
    \centerline{\hspace{1.cm}\epsfxsize=3.7cm\epsfbox{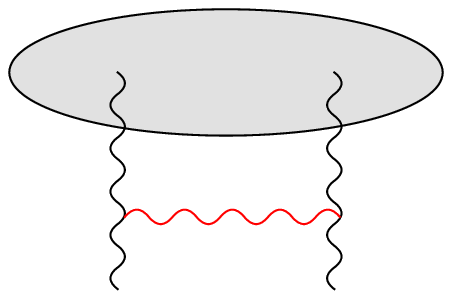}
    \hspace{.3cm}\epsfxsize=4.cm\epsfbox{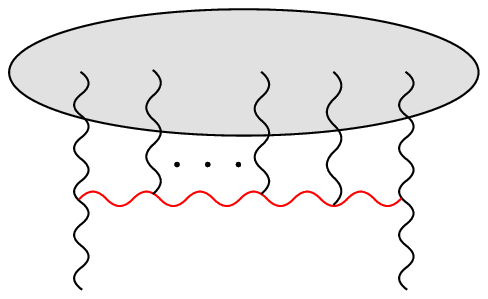}
    \hspace{.3cm}\epsfxsize=4.cm\epsfbox{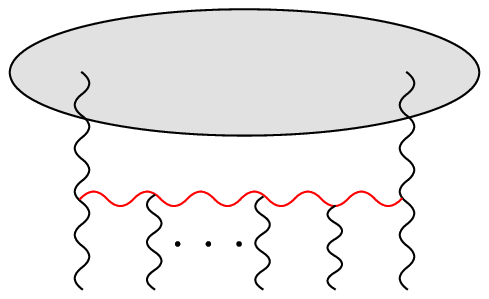}}
 \vspace*{0.2cm}
 \hspace{3.7cm} (a)\hspace{3.7cm} (b)\hspace{4.cm}(c)
    \caption{\sl Gluon processes which occur
    in one step of high energy
    evolution. \label{BREMfig}}
\end{figure}

In turn, the small--$x$ gluons are produced through quantum
evolution, i.e., through radiation from {\em color sources}
(typically, other gluons) with larger values of $x$, whose
internal dynamics is `frozen' by Lorentz time dilation. Let $\tau
=\ln 1/x$ denote the {\em rapidity\,}; it takes, roughly, a
rapidity interval $\Delta\tau \sim 1/\alpha_s$ to emit one
small--$x$ gluon; thus, in the high energy regime where
$\alpha_s\tau \gg 1$, the dipole meets with well developed gluon
cascades, as illustrated  in Fig. \ref{dis_blois1}. Three types of
processes can be distinguished in Fig. \ref{dis_blois1}, which for
more clarity are singled out in Fig. \ref{BREMfig}.

The first process, Fig. \ref{BREMfig}.a, represents one step in
the standard BFKL evolution \cite{CGC}; by iterating this step,
one generates gluon ladders which are resummed in the solution to
the BFKL equation \cite{CGC}. However, by itself, the latter is
well known to suffer from conceptual difficulties in the high
energy limit : \texttt{(i)} The BFKL estimate for the dipole
scattering amplitude $T_\tau(r)$ grows exponentially with $\tau$
(i.e., like a power of the energy), and thus eventually violates
the unitarity bound $T_\tau(r)\le 1$. (The upper limit $T_\tau=1$
corresponds to the black disk limit, in which the dipole is
totally absorbed by the target.) \texttt{(ii)} The BFKL ladder is
not protected from deviations towards the non--perturbative domain
at low transverse momenta $k_\perp^2\simle \Lam^2$ (`infrared
diffusion'). With increasing energy, the BFKL solution receives
larger and larger contributions from such soft intermediate
gluons, and thus becomes unreliable.

These `small--$x$ problems' of the BFKL equation are both cured by
the {\em gluon recombination} processes $(n\to 2)$ illustrated in
Fig. \ref{BREMfig}.b which are important at high energy, when the
gluon density in the target is large, and lead to {\em gluon
saturation} and the formation of the CGC. Such processes are
included in the Balitsky--JIMWLK equation \cite{CGC}, a
non--linear, functional, generalization of the BFKL evolution
which describes the approach towards gluon saturation in the
target and preserves the unitarity bound in the evolution of the
scattering amplitudes.

However, the Balitsky--JIMWLK equation misses \cite{IT04} the
process in Fig. \ref{BREMfig}.c --- the  $2\to n$ gluon splitting
--- which describes the bremsstrahlung of additional small--$x$
gluons in one step of the evolution. By itself, this process is
important in the  {\em dilute} regime, where it leads to the
construction of higher--point gluon correlation functions from the
dominant 2--point function. But once generated, the $n$--point
functions with $n>2$ are rapidly amplified by the subsequent BFKL
evolution (the faster the larger is $n$) and then they play an
important role in the non--linear dynamics leading to saturation.
Thus, such splitting processes {\em are} in fact important for the
evolution towards high gluon density, as originally observed in
numerical simulations \cite{Salam} of Mueller's `color dipole
picture' \cite{Mueller}, and more recently reiterated in the
analysis in Refs. \cite{MS04,IMM04,IT04}.

Equations including both merging and splitting in the limit where
the number of colors $N_c$ is large have recently became available
\cite{IT04} (see also Refs. \cite{MSW05,BIIT05}), but their
general solutions have not yet been investigated (except under
some additional approximations \cite{IT04,Soyez}). Still, as we
shall argue now, the {\em asymptotic} behaviour of the
corresponding solutions --- where by `asymptotic' we mean both the
high--energy limit $\tau\to\infty$ and the weak coupling limit
$\alpha_s\to 0$ --- can be {\em a priori} deduced from
universality considerations, by exploiting the correspondence
between high--energy QCD and the reaction--diffusion problem of
statistical physics \cite{IMM04}.

To that aim, it is convenient to rely on the event--by--event
description \cite{IMM04} of the scattering between the external
dipole and the hadronic target (cf. Fig. \ref{dis_blois1}) and to
use the large--$N_c$ approximation to replace the gluons in the
target wavefunction by color dipoles 
\cite{Mueller}. Then, the dipole--target scattering amplitude
corresponding to a given event can be estimated as \be\label{Tf}
  T_{\tau}(r,b)
  \,\simeq \,\alpha_s^2 \,f_{\tau} (r,b)\,,\ee
where $\alpha_s^2$ is the scattering amplitude between two dipoles
with comparable sizes and nearby impact parameters, and $f_{\tau}
(r,b)$ is the {\em occupation number} for target dipoles with size
$r$ at impact parameter $b$, and is a {\em discrete} quantity:
$f=0,1,2,\dots$. Thus, in a given event, the scattering amplitude
is a multiple integer of $\alpha_s^2$.

In this dipole language, the $2\to 4$ gluon splitting depicted in
Fig. \ref{BREMfig}.c is tantamount to $1\to 2$ dipole splitting,
and generates {\em fluctuations} in the dipole occupation number
and hence in the scattering amplitude. Thus, the evolution of the
amplitude $T_{\tau}(r,b)$ with increasing $\tau$ represents a {\em
stochastic process} characterized by an expectation value $\lan
T(r,b)\ran_{\tau} \simeq \alpha_s^2 \,\lan f (r,b) \ran_{\tau}$,
and also by fluctuations $\delta T \sim \alpha_s^2\delta f \sim
\sqrt{\alpha_s^2 T}$ (we have used the fact that $\delta f \sim
\sqrt{f}$ for fluctuations in the particle number). Clearly, such
fluctuations are relatively important (in the sense that $\delta T
\simge  T$) only in the {\em very} dilute regime where $\lan f
\ran\simle 1$, or $\lan T\ran\simle \alpha_s^2$.

Eq.~(\ref{Tf}) applies so long as the scattering is weak, $ T \ll
1$, but by extrapolation it shows that the unitarity corrections
are expected to be important when the dipole occupation factor
becomes of order $1/\alpha_s^2$. Consider first the formal limit
$\alpha_s^2\to 0$, in which the maximal occupation number $N\sim
1/\alpha_s^2$ becomes arbitrarily large. Then one can neglect the
particle number fluctuations and follow the evolution of the
scattering amplitude in the {\em mean field approximation} (MFA).
This is described by the Balitsky--Kovchegov equation \cite{CGC},
a non--linear version of the BFKL equation which, as shown in Ref.
\cite{MP03}, lies in the same universality class as the
Fisher--Kolmogorov--Petrovsky--Piscounov (FKPP) equation
 (the MFA for the reaction--diffusion process and
related phenomena in biology, chemistry, astrophysics, etc; see
\cite{Saar} for recent reviews and more references). The FKPP
equation reads, schematically,
 \be\label{BK}
 \partial_\tau T(\rho,\tau)\,=\,
 \partial_\rho^2 T(\rho,\tau)\,+\,
 T(\rho,\tau)\,- \,T^2(\rho,\tau),\ee
in notations appropriate for the dipole scattering problem:
$T(\rho,\tau)\equiv \lan T(r)\ran_{\tau}$ and $\rho\equiv
\ln(r_0^2/r^2)$, with $r_0$ a scale introduced by the initial
conditions at low energy. Note that weak scattering ($T\ll 1$)
corresponds to small dipole sizes ($r\ll 1/Q_s$), and thus to
large values of $\rho$.
The three terms on the r.h.s. of Eq.~(\ref{BK}) describe,
respectively, diffusion, growth and recombination. The first two
among them represent (an approximate version of) the BFKL
dynamics, while the latter is the non--linear term which describes
multiple scattering and thus ensures that the evolution is
consistent with the unitarity bound $T\le 1$.

\begin{figure}[t]    
    \centerline{\epsfxsize=15.cm\epsfbox{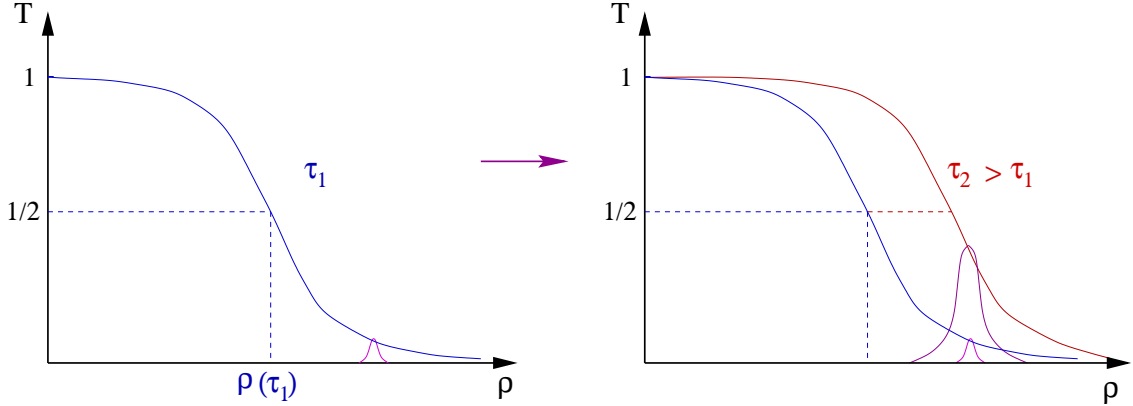}}
    \caption{\sl Evolution of the continuum front of the Balitsky--Kovchegov
    equation with increasing $\tau$.
                 \label{TWave5}}
\end{figure}

Specifically, the solution $T_\tau(\rho)$ to Eq.~(\ref{BK}) is a
{\em front} which interpolates between two fixed points : the
saturation fixed point $T=1$ at $\rho\to -\infty$ and the unstable
fixed point $T=0$ at $\rho\to \infty$ (see Fig. \ref{TWave5}). The
position of the front, which marks the transition between strong
scattering ($T\sim 1$) and, respectively, weak scattering ($T\ll
1$), defines the {\em saturation scale\,}: $\rho_s(\tau)\equiv
\ln(r_0^2 Q_s^2(\tau))$. With increasing $\tau$, the front moves
towards larger values of $\rho$, as illustrated in Fig.
\ref{TWave5}. Note that the dominant mechanism for propagation is
the BFKL growth in the tail of the distribution at large $\rho$ :
the front is {\em pulled} by the rapid growth of a small
perturbation around the unstable state. In view of that, the {\em
velocity} of the front $\lambda\equiv {d\rho_s}/{d\tau}$ is fully
determined by the linearized version of Eq.~(\ref{BK}), which
describes the dynamics in the tail. Specifically, by solving the
BFKL equation one finds \cite{SCALING,MT02,MP03} that, for $\rho
>\rho_s(\tau)$ and sufficiently large $\tau$,
 \be \label{TBFKL}
 T_\tau(\rho) \,\simeq\,{\rm e}^{\omega \bar\alpha_s \tau}
 \,{\rm e}^{-\gamma\rho}  \,=\,{\rm e}^{-\gamma(\rho -\rho_s(\tau))}\,
 ,\qquad \rho_s(\tau)\equiv c \bar\alpha_s \tau,\ee
where $\bar{\alpha}_s = {\alpha}_s N_c/\pi$, $\gamma=0.63..$, and
$c \equiv \omega/\gamma=4.88..\,$. From Eq.~(\ref{TBFKL}) one can
immediately identify the velocity of the front in the MFA as
$\lambda_0 = c\bar\alpha_s$. Since $Q_s^2(\tau) \simeq Q_0^2\,
{\rm e}^{\lambda_0 \tau}$, it is furthermore clear that
$\lambda_0$ plays also the role of the {\em saturation exponent}
(here, in the MFA).

\comment{ Note also that the propagation of the front, as
described by Eq.~(\ref{TBFKL}), represents a {\em traveling wave}
\cite{MP03} : in a comoving frame, the shape of the front is
independent of $\tau$. This is the property originally referred to
as `geometric scaling' \cite{geometric,SCALING}, and which might
explain a remarkable regularity observed in the small--$x$ data
for DIS at HERA \cite{geometric}. Namely, for $x\le 0.01$, the
total cross--section $\sigma_{\gamma^*p}(x,Q^2)$ for the
absorbtion of the virtual photon shows approximate scaling as a
function of $Q^2/Q_s^2(x)$, with $Q_s^2(x) \propto (1/x)^\lambda$
and $\lambda \approx 0.3$ from a fit to the data.}

\begin{figure}[t]    
    \centerline{\epsfxsize=14.cm\epsfbox{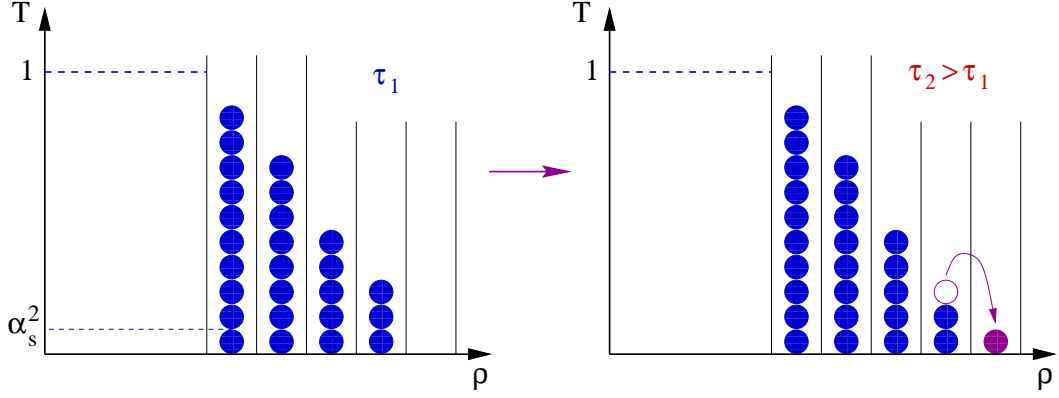}}
    \caption{\sl Evolution of the discrete front of a microscopic
    event with increasing rapidity $\tau$. The small blobs
    are meant to represent the elementary
    quanta $\alpha_s^2$ of $T$ in a microscopic event.
                 \label{TWave6}}
\end{figure}

What is the validity of the mean field approximation ? We have
earlier argued that the gluon splitting processes (cf. Fig.
\ref{BREMfig}.c) responsible for dipole number fluctuations should
play an important role in the dilute regime. This is further
supported by the above considerations on the {\em pulled} nature
of the front: Since the propagation of the front is driven by the
dynamics in its tail where the fluctuations are {\em a priori}
important, the front properties should be strongly sensitive to
fluctuations. This is indeed known to be the case for the
corresponding problem in statistical physics \cite{BD,Saar}, as it
can be understood from the following, qualitative, argument:

Consider a particular realization of the stochastic evolution of
the target, and the corresponding scattering amplitude which is
discrete (in steps of $\alpha_s^2$). Because of discreteness, the
microscopic front looks like a histogram and thus is necessarily
{\em compact} : for any $\tau$, there is only a finite number of
bins in $\rho$ ahead of $\rho_s$ where $T_\tau$ is non--zero (see
Fig. \ref{TWave6}). This property has important consequences for
the propagation of the front. In the empty bins on the right of
the tip of the front, the local, BFKL--like, growth is not
possible anymore (as this would require a seed). Thus, the only
way for the front to progress there is via {\it diffusion}, i.e.,
via radiation from the occupied bins at $\rho <\rho_{\rm tip}$
(compare in that respect Figs. \ref{TWave5} and \ref{TWave6}). But
since diffusion is less effective than the local growth, we expect
the velocity of the microscopic front (i.e., the saturation
exponent) to be reduced as compared to the respective prediction
of the MFA.

To obtain an estimate for this effect, we shall rely again on the
universality of the asymptotic ($\tau\to\infty$ and $N\equiv
1/\alpha_s^2\gg 1$) behaviour\cite{IMM04}. Namely, from the
experience with the reaction--diffusion process and related
problems in statistical physics \cite{BD,Saar}, one knows indeed
that the dominant behaviour for large evolution `time' and large
(but finite) occupancy $N\gg 1$ is independent of the details of
the microscopic dynamics, and thus is the same for all the
processes whose mean field limit ($N\to\infty$) is governed by the
FKPP equation (\ref{BK}). In particular, the dominant contribution
to the correction $\lambda_N- \lambda_0$ to the front velocity is
known to be universal, and can be obtained through the following,
intuitive, argument, due to Brunet and Derrida \cite{BD}:

For a given microscopic front and $N\gg 1$, the MFA should work
reasonably well everywhere except in the vicinity of the tip of
the front, where the occupation number $f$ becomes of order one
(corresponding to $T\sim \alpha_s^2$ in the QCD problem) and thus
the linear growth term becomes ineffective.
Accordingly, Brunet and Derrida suggested a modified version of
the FKPP equation (\ref{BK}) in which the `BFKL--like' growth term
is switched off when $T< \alpha_s^2$ :
 \be\label{BKBD}
 \partial_\tau T(\rho,\tau)\,=\,
 \partial_\rho^2 T\,+\,\Theta\big(T - \alpha_s^2\big)
 T(1- T).\ee
By solving this equation in the linear regime, they have obtained
the first correction to the front velocity as compared to the MFA
(in notations adapted to QCD; see Ref. \cite{IMM04} for details):
 \be\label{ls}
 \lambda\,\simeq\,\bar\alpha_s\left[c\,-\,
 \frac{\kappa}{\ln^2(1/\alpha_s^2)}\,+\,{\cal O}
 \big(1/\ln^3 \alpha_s^2\big)\right]
 \,,
 \ee
where the numbers $c \approx 4.88$ and $\kappa \approx 150$ are
fully determined by the linear (BFKL) equation. In QCD, the same
result has been first obtained through a different but related
argument by Mueller and Shoshi \cite{MS04}. Note the extremely
slow convergence of this result towards its mean field limit: the
corrective term vanishes only logarithmically with decreasing
$1/N=\alpha_s^2$, rather than the power--like suppression usually
found for the effects of fluctuations. This is related to the high
sensitivity of the pulled fronts to fluctuations, as alluded to
above. Such a slow convergence, together with the relatively large
value of the numerical factor  $\kappa$, make that the above
estimate for $\lambda$, although {\em exact} for asymptotically
small  $\alpha_s^2$, is in fact useless for practical
applications. To understand the subasymptotic corrections and,
more generally, the behaviour of the saturation momentum and of
the scattering amplitudes for realistic values of $\tau$ and
$\alpha_s$, one needs to solve the actual evolution equations of
QCD \cite{IT04,MSW05,BIIT05}, a program which is currently under
way \cite{Soyez}.

\end{document}